\documentclass[12pt]{article}
\usepackage{cite}
\usepackage{avant}
\usepackage{amssymb}
\usepackage{multirow,booktabs,supertabular,multirow}
\usepackage{pifont}\usepackage{amsmath}
\numberwithin{equation}{section}
\newcommand\email[4]{#1@#2.#3.#4}

\newcommand\Adj[1]{\operatorname{Adj}({#1})}
\def\C{{\mathbf{C}}}

\newcommand\ch{\operatorname{ch}}

\newcommand\Ext{\operatorname{Ext}}

\def\eq#1{(\ref{#1})}

\def\T{{T}}

\def\tr{\operatorname{Tr}}

\def\Z{{Z}}
\newfont\sheafnt{rsfs10}

\newcommand\npb[3]{Nucl.~Phys. {\bf B{#1}}~({#2})~{#3}}

\begin{document}
\title{Forbidden territories in the string landscape}
\author{
Alok Kumar\thanks{\email{kumar}{iopb}{res}{in}} \\
\small Institute of Physics, Bhubaneswar 751~005, India.
\and Subir Mukhopadhyay
\thanks{\email{subir}{iopb}{res}{in}. (On leave from IOP, Bhubaneswar)} \\
\small Institute for Studies in Theoretical Physics and
Mathematics,\\\small P.O. Box 19395-5531, Tehran, Iran \and
Koushik Ray \thanks{\email{koushik}{iacs}{res}{in}}\\
\small Department of Theoretical Physics \& Centre for Theoretical Sciences\\
\small  Indian Association for
 the Cultivation of Science \\
\small  Calcutta 700 032, India.
}
\date{}
\maketitle
\begin{abstract}
\small
\noindent
Problems of stabilizing moduli of the type--IIB string theory on 
toroidal orientifolds $\T^6/\Z_2$, in presence of worldvolume fluxes 
on various D-branes, are considered. For $Z_2$ actions, introducing
either O9 or O3 planes, we rule out the possibility of moduli stabilization 
in a wide class of models with $\mathcal{N}=1$ supersymmetry, characterized by
the type of fluxes turned on along D-brane worldvolume. Our results, in particular,
imply that Abelian worldvolume fluxes can not by themselves  
stabilize closed string moduli, in a consistent supersymmtric model,
for above orientifold compactifications. We also discuss other 
$Z_2$ orientifolds of $T^6$ and show that certain other brane wrappings 
are also ruled out by similar consistency requirements. 
In specific setups we consider examples with 
D9-branes wrapping on a complex three-torus with its world-volume fluxes taken to
be  semi-homogeneous bundles and D7-branes wrapping
holomorphic four-cycles of the complex three-torus carrying world-volume fluxes.
\end{abstract}
\thispagestyle{empty}
\clearpage
\section{Introduction}
In a world abound with three-form fluxes\cite{gkp,gp} in the bulk
and gauge fluxes\cite{as,mmms,bul1} on D-branes, assuming a copious
supply of all kinds of necessary fluxes and branes, many of the
closed-string moduli arising from  compactification can be fixed
\cite{kst,Review,AM,Blumenhagen,Cascales2,Bianchi1,akm,kmr}. Among
these the complex-structure and axion-dilaton moduli of the
type--IIB string theory are wont to be fixed by three-form NS-NS and
R-R fluxes, preserving an $\mathcal{N}=1$ supersymmetry. In this
framework stabilizing the K\"ahler moduli, the overall volume of the
compact target in particular, turns out to be a bit of a
contretemps. The mechanisms envisaged hitherto for achieving such a
lofty goal rely on non-perturbative means of generating
superpotentials through gaugino condensations. On the other hand,
fluxes associated with gauge fields on the brane generate D-term
potential and stabilize K\"ahler moduli. However, it has been shown
recently \cite{akm2} that to build a consistent model of this type
for toroidal orientifolds, turning on vacuum expectation values
(VEV) for the scalars charged under the associated $U(1)$ symmetry
is required in addition. The validity of these results is, however,
restricted to instances of small VEV for the scalars. Only a small
patch of the open string moduli space may thus be explored in this
scheme and the method has little to say about the vast string
landscape. With several schemes proposed to stabilize the plethora
of moduli and yet our own world looming heavy upon us, awaiting to
be ``formulated'' by string theory, amidst ``hundred indecisions"
and ``hundred visions and revisions", it is important to narrow down
the possibilities by eliminating at least parts of such schemes by
their internal consistency.

In this context, a standard lore has been that Abelian gauge fluxes
on the D-brane worldvolume can not stabilize closed string
moduli, consistent with $\mathcal{N}=1$ supersymmetry and tadpole cancelations,
in toroidal orientifold compactifications of IIB string theory.
The reason behind such a belief is that fluxes generate
new tadpoles which can not be cancelled unless extra orientifold 
planes are also introduced. However, such O-planes are possible
to introduce only in the case of further orbifoldings, a process which
introduces other complications from the stabilization point of view,
such as the presence of twisted sector moduli. In this article, however,
we restrict to the issue of closed string moduli stabilization in 
toroidal orientifolds only. In particular, we consider four different
IIB string orientifolds of a six-torus, that is,
$\T^6/\Z_2$, with D-branes carrying magnetic fluxes on their
world-volume wrapped on holomorphic cycles of the six-torus.
We denote these four different orientifolding actions by
$Z_2^A$, $Z_2^B$, $Z_2^C$ and $Z_2^D$ respectively.
The first one is given by $Z_2 \equiv \Omega \equiv Z_2^A$ and the construction  also corresponds to 
a type I compactification on $T^6$. The second one is 
$Z_2 \equiv \Omega (-)^{F_L} I_6 \equiv Z_2^B$ orientifold, with $I_6$ being 
the inversion on six internal coordinates, which has been discussed a lot in the context of closed string flux 
compactification\cite{gkp,gp,kst}.  

In this paper, we are able to rule out the possibilities of consistent
compactifications on $T^6$ tori with $Z_2^A$ and $Z_2^B$
orienfoldings, with known holomorphic vector bundles on available
D-branes. Though we consider more general situations, however, 
our results imply that Abelian gauge fluxes, in particular, can not be used 
for moduli stabilizations in the above contexts, thus confirming the
general conjecture in a rigorous fashion.

We also present the basic setup to analyze the possibilities
of moduli stabilizations in consistent toroidal orientifolds of 
$T^6$ with the other two orientifolding actions given by:
$Z_2 \equiv \Omega (-)^{F_L} I_2 \equiv Z_2^C$ and 
$Z_2 \equiv \Omega I_4 \equiv Z_2^D$. Though in these cases,
a nogo result is harder to obtain due to the particular 
inhomogeneous structure of the tadpole constraints, nevertheless
we are able to rule out certain specific flux brane configurations
in a supersymmetric setup.

We consider two nontrivial classes of models. In the first, space-filling 
magnetized D9-branes are wrapped on the six-torus itself. In the other 
space-filling D7-branes wrapped on different holomorphic four-cycles of the
six-torus are considered. In the case of the third possibility, 
namely the use of magnetized  D5-branes, 
the relevant arguments are presented in
section-\ref{D5-branes}. The fluxes on the world-volume of D-branes
are taken to be non-Abelian (also used earlier in different contexts
\cite{blu2}) in general and we draw our conclusions by analyzing 
the consistency conditions in presence of various orientifold planes 
arising from different  orientifolding actions.

The scope of our approach is
limited by the incompleteness of the classification of vector
bundles on tori. In case of D9-branes we consider semi-homogeneous
vector bundles on a complex three-torus and show that such bundles
satisfying all the requirements fail to exist. This, in particular,
confirms the suspicion that there is not even an Abelian
configuration which may achieve the cherished goal. In the
cases with D7-branes, too, we show that a gauge theoretic configuration satisfying
all the requirements of supersymmetry as well as tadpole cancelation
is an impossibility. 
The rest of the paper is organized as following: in section-2 we present
the supersymmetry and tadpole constraints for the D9 brane system with 
semi-homogeneous vector bundles and analyze those constraints to show
our nogo theorem for $Z_2^A$ and $Z_2^B$ orientifolds, 
giving rise to O9 and O3 planes respectively. In section-3, a
similar exercise is repeated for the magnetized D7-branes. In this
case, the use of O9- corresponding to $Z_2^A$ orientifolding 
is trivially ruled out. In addition, we also rule out O3 and 
O7 possibilities corresponding to $Z_2^C$ and $Z_2^D$ respectively.
In subsetions-\ref{D5-branes}, we also point out that D5 branes can not
be used for moduli stabilizations in $Z_2^A$, $Z_2^B$ and $Z_2^C$
examples given above.
In section-4 we collect all the results, to prove  
that closed string moduli can not be consistently stabilized in 
$Z_2^A$ and $Z_2^B$ orientifold compactification on $T^6$
by the choice of gauge fluxes used thus far, including close 
string fluxes.

\section{Nine-branes} \label{Nine-branes}
First, we consider space-filling magnetized D9-branes, that is, 
D9-branes carrying constant magnetic fluxes on their
world-volumes, wrapped on the orientifolded $\T^6$ along
with orientifold three-planes \cite{AM,akm,akm2}.
\subsection{Consistency requirements}
We shall look upon the $\T^6$ as a complex three-torus or a
three-dimensional compact complex Abelian variety and denote it by
$X$. An ${\mathcal N}=1$ supersymmetric configuration of $N$ number
of magnetized D9-branes wrapped on $X$ is given by a vector bundle
on $X$, $\oplus_{k=1}^N E_k$, where each addendum corresponds to a
stack of branes. In the presence of an orientifold plane, in order for the
orientifold plane and all the branes to preserve the same
supersymmetry, the central charge
\begin{equation}
\label{cc1}
{\mathcal Z}(E_k)= \int_X e^{-i\Omega}\ch(E_k)
\end{equation}
of the branes must be such that
\begin{equation}
\label{susy}
\mathrm{Im}~ (e^{-i\theta}{\mathcal Z}(E_k)) = 0,\qquad \mathrm{Re}~(e^{-i\theta}{\mathcal Z}(E_k)) <0
\end{equation}
for all $k$ for a certain K\"ahler 2-form $\Omega$ on $X$. Here
$\ch(E)$ denotes the Chern character of $E$ and value of $\theta$ depends on the orientifold plane\cite{AM}. The D brane tadpoles 
in this notation, which corresponds to \cite{kmr}, can be
obtained from the Wess-Zumino action, depends on the choice of the orientifolding. For $Z_2^A$ and $Z_2^B$ respectively 
the WZ actions read:
\begin{gather}
V_{\text wz} = \sum\limits_{k=1}^{N} \int\limits_{M^k_{10}} [
\ch_1(E_k)\wedge C_8 + \ch_3(E_k) \wedge C_4 ],\label{csA}\\
V_{\text wz} = \sum\limits_{k=1}^{N} \int\limits_{M^k_{10}} [
 \ch_0(E_k)\wedge C_{10} + \ch_2(E_k) \wedge C_6 ],\label{csB}
\end{gather}
where we write $M^k_{10}$ for the world-volume of the D9-brane in
the $k$-th stack. We will discuss the tadpoles arising for different choices or orientifolding. The tadpoles, along with the contribution of orientifold plane, when added up over all the stacks should vanish.
A valid model with moduli stabilized is a solution to the
constraints \eq{susy} and satisfies the tadpole cancelation
condition. Indeed, the existence of a Hermitian two-form $\Omega$ as
a solution to \eq{susy}, with a given set of vector bundles on $X$,
consistent with vanishing tadpole, goes by the name of stabilization
of K\"ahler moduli. We shall consider semi-homogeneous vector
bundles $E_k$, of ranks $r_k$, respectively, on $X$.
\subsection{Semi-homogeneous bundles} \label{Semi-homogeneous bundles}
Semi-homogeneous vector bundles on complex tori have been classified.
Let $X$ be an $n$-dimensional
Abelian variety over $\C$ and $E$ a vector bundle of rank $r$ over $X$.
Then $E$ is called \emph{semi-homogeneous} if $\dim\Ext^1(E,E)=n$.
Semi-homogeneous bundles are Gieseker semi-stable.
The Chern character of $E$ assumes the form \cite{mukai,maciocia}
\begin{equation}
\ch(E)=( r, c, c^2/2r, \cdots c^n/n!\,r^{n-1}).
\end{equation}

Let us now consider the bundle $\oplus_k E_k$ on $X$. Let us denote
the rank of the addendum $E_k$ by $r_k$. Denoting by $H_k$ the
Hermitian matrix corresponding to the first Chern class $c_1(E_k)$
of $E_k$, each of $H_k$ is a $3\times 3$ Hermitian matrix. We shall
denote the $3\times 3$ non-singular Hermitian matrix corresponding
to the K\"ahler form by $\Omega$ too. 
\subsection{The constraints}
In  this subsection we explicitly lay down the supersymmetry \eq{susy}
and tadpole cancellation conditions for different orientifold planes
using notations introduced in the last subsection. 
In order to keep the discussion simple, we only
elaborate, wherever possible, on situations with positive wrapping 
numbers, defined by the the Jacobian of the matrices mapping the
worlvolume of the brane to embedding space. 
First we starting by 
setting the wrapping matrices to identity for 
all the stacks. For O5-plane, however we need to introduce negative 
wrapping number as well. We emphasize, however, that 
our analysis is valid for any general wrapping as we discuss later.
\subsubsection*{O3-plane}
Let us begin with the O3-plane in a type--IIB compactification
$T^6/Z_2^B$, with $Z_2^B$ defined earlier. It turns out, that in presence of O3-plane either all stacks have $\theta=0$ or all stacks have $\theta=\pi$. For $\theta=0$
preservation of $\mathcal{N}=1$ supersymmetry requires,
by \eq{susy}, 
\begin{gather}
\det\Omega - \frac{1}{r_k} \tr(\Omega \Adj{H_k})=0,\label{suseo31}\\
\tr(H_k \Adj{\Omega})< \frac{1}{r_k^2}\det H_k, \label{suseo32}
\end{gather}
for each stack indexed by $k$.
For $\theta=\pi$ \eq{suseo31} remains the same but the inequality sign 
\begin{gather}
\det \Omega - \frac{1}{r_k} \tr(\Omega \Adj{H_k})=0,\quad \forall k\label{suseo71}\\
\tr(H_k \Adj{\Omega})> \frac{1}{r_k^2}\det H_k, \quad\forall k .\label{suseo72}
\end{gather}
For an O3-plane we only need consider the D7-brane and D3-brane tadpoles 
whose cancellation leads to the equations, 
\begin{gather}
\sum_k H_k=0,\label{o3tadp1}\\
\sum_k\frac{1}{r_k^2}\det H_k\leq 16\label{o3tadp2},
\end{gather}
respectively.
Let us note that since the O3-plane is transeverse to the compact
space $T^6$, 
the supersymmetry and tadpole cancellation conditions are 
invariant under rotations of the torus.
\subsubsection*{O7-plane}
The supersymmetry conditions in such a compactification, 
on orientifolds $T^6/Z_2^C$,
remains the same as that of O3-plane as
the allowed values are once again $\theta=0$ or $\theta=\pi$. 
So preservation of supersymmetry requires either (\ref{suseo31}, \ref{suseo32}) or (\ref{suseo71}, \ref{suseo72}).
The tadpole condition for O7-plane is different and depends on the choice of the orientifolding action. Without any loss of generality we choose the orientifolding action acting on complex coordinates as $(z^1, {\bar z}^1) \longrightarrow (-z^1, -{\bar z}^1)$ while keeping the rest fixed. In this case, 
the tadpole condition becomes 
\begin{gather}
\sum_k H^{(1{\bar 1})}_k \leq 16,
\quad
\sum_k H^{(i{\bar j})}_k =0,
\quad (i{\bar j})\neq (1{\bar 1})
\label{o7tadp1}\\
\sum_k\frac{1}{r_k^2}\det H_k=0\label{o7tadp2}.
\end{gather}
Here the tadpole cancellation conditions are not invariant under rotation along $T^6$ directions.

\subsubsection*{O9-plane}
Attempts of stabilizing moduli with
stacks of D9-branes in the presence of O9-planes 
have been made earlier \cite{AM} along with Abelian fluxes. 
None of the various examples, however, satisfied \emph{all} the consistency
conditions. Preservation of an $\mathcal{N}=1$ supersymmetry of this
instance, which corresponds to the orientifolding $T^6/Z_2^D$,
requires $\theta=-\frac{\pi}{2}$. That implies following constraints:
\begin{gather}
\det \Omega - \frac{1}{r_k} \tr(\Omega \Adj{H_k})>0,\quad \forall k\label{suseo91}\\
\tr(H_k \Adj{\Omega})= \frac{1}{r_k^2}\det H_k, \quad\forall k\label{suseo92}.
\end{gather}
In the presence of an O9-plane we need to consider the cancellation of
D5- and D9-brane tadpoles, leading to
\begin{gather} 
\sum_k\frac{1}{r_k^2}\det H_k = 0 \label{o9tadp1},\\
\sum_k r_k \leq 16 \label{o9tadp2}.
\end{gather}

\subsubsection*{O5-plane}
For this case, on orientifold: $T^6/Z_2^D$, the supersymmetry condition requires either $\theta=\frac{\pi}{2}$ or $\theta=-\frac{\pi}{2}$.
The first condition corresponds to that of the O9-plane, 
and are given by (\ref{suseo91}, \ref{suseo92}).
For $\theta=-\frac{\pi}{2}$ the conditions become:
\begin{gather}
 \det \Omega - \frac{1}{r_k} \tr(\Omega \Adj{H_k})  < 0,\quad \forall k ,\label{suseo51}\\
\tr(H_k \Adj{\Omega})= \frac{1}{r_k^2}\det H_k, \quad\forall k .\label{suseo52}
\end{gather}
The tadpole condition for O5-plane, like O7-plane depends on the choice of the 
orientifolding action. This time we choose the orientifolding action acting on complex 
coordinates as $(z^2, z^3, {\bar z}^2, 
{\bar z}^3) \longrightarrow (-z^2, -z^3, -{\bar z}^2, -{\bar z}^3)$ 
while keeping the rest fixed. The tadpole condition becomes 
\begin{gather}
\sum_k n_k ~ r_k =0,
\label{o5tadp1}\\
\sum_k n_k ~ \Adj{H_k}^{(1{\bar 1})} \leq 16,
\quad\quad
\sum_k n_k ~ \Adj{H_k}^{(i{\bar j})} = 0, \quad \forall (i{\bar j}) \neq (1{\bar 1})
\label{o5tadp2}.
\end{gather}
In this case we have to allow the overall wrapping number $n_k$ so that it can take both positive and negative values. Once again, the tadpole cancellation conditions are not invariant under rotation along $T^6$.

In the next subsection we present a general analysis of the supersymmetry and tadpole 
cancellation conditions for the cases of O3 and O9-planes and discuss the cases of 
O5 and O7-planes briefly. 

\subsection{Analysis of constraints}
In this subsection we will analyze the mutual consistency of  the constraints 
for two cases, namely: O3 and O9. That takes care of all the $Z_2$ orientifolding 
actions with fixed 3-planes and 9-planes. The other orientifolding actions, 
for which one has fixed 5-plane or 7-plane the tadpole cancelation conditions do not have homogeneous forms and are not invariant under rotation among the directions of $T^6$. So these cases do not admit a general analysis and 
require a case by case discussion. 
Nevertheless, as emphasized earlier, our results in this section as well as in the later
ones, are sufficient to prove that there is no consistent toroidal orientifold 
of $T^6$ with either O3 or  O9 planes, within the class of worldvolume fluxes that
are being considered.

We begin with \eq{suseo31}--\eq{o3tadp2} and \eq{suseo91}--\eq{o9tadp2} for a set 
of Hermitian matrices $\Omega$,
$H_k$, $k=1, \ldots , N$ for an arbitrary fixed positive integer,
$N$. We are, therefore, considering $N$ stacks of D-branes on the
three-torus, each stack $k$ corresponding to a semi-homogeneous
vector bundle $E_k$ of rank $r_k$, with first Chern class given by
the Hermitian matrix $H_k$. Clearly, for every $k$, the rank $r_k$
of the bundle $E_k$ is a non-zero positive integer. First, let us
consider the case in which all the matrices $H_k$ are non-singular.
In this cases it turns out to be convenient to define a new non-singular matrix $G_k=H_k \Omega^{-1}$ and use
the relation $\Adj{M}=(\det{M})\,M^{-1}$ for a non-singular matrix
$M$. In this notation the constraints arising from supersymmetry and 
tadpole cancellations are as follows.

\subsubsection*{O3-plane} 
To begin with we consider $\theta=0$.
We rewrite the relations \eq{suseo31}, \eq{suseo32} and \eq{o3tadp1}
respectively as
\begin{gather}
r_k= \tr ({G_k^{-1}})\det{G_k}, \label{go31}\\
\tr(G_k)< \frac{1}{r_k^2}\det {G_k},\label{go32}\\
\sum_k G_k=0, \label{o3desum}
\end{gather}
where we have used the strict positivity of $r_k$ and $\det\Omega$.

Since the first two equations involve only traces and determinants of $G_k$, we can
rewrite them in terms of the eigenvalues of the matrices.
Denoting the eigenvalues of the $3\times 3$ matrix $G_k$ as $x_k,y_k,z_k$,
we rewrite equations \eq{go31} and \eq{go32} as, respectively,
\begin{gather}
r_k = x_k y_k+y_k z_k+ z_k x_k, \label{eigeno31}\\
r_k^2(x_k+y_k+z_k) < x_k y_k z_k.\label{eigeno32}
\end{gather}
Four possibilities arise for the combination of
signs of the eigenvalues $x_k, y_k, z_k$. Let us now discuss them in turn.
\begin{enumerate}
\item \label{cas1}
All the eigenvalues are positive.
In this case, \eq{eigeno31} and \eq{eigeno32} together imply
\begin{equation*}
r_k (x_k y_k+y_k z_k+ z_k x_k)(x_k+y_k+z_k) < x_k y_k z_k,
\end{equation*}
as $r_k = x_k y_k+y_k z_k+ z_k x_k>0$.
Using the inequality,
\begin{equation*}
\begin{split}
(x_k y_k+y_k z_k+ z_k x_k)(x_k+y_k+z_k) &-9x_k y_k z_k \\&=
x_k(y_k-z_k)^2+y_k(z_k-x_k)^2+z_k(x_k-y_k)^2\\
& > 0,
\end{split}
\end{equation*}
for non-vanishing positive numbers, we thus require
\begin{equation*}
r_k < {1}/{9}.
\end{equation*}
Since $r_k$ is a non-zero positive integer, this is not possible.
\item \label{cas2}
Two of the eigenvalues are positive and one negative.
Let us consider a particular value of $k$ and,
without loss of generality, let us take $x_k=a$, $y_k=b$ and $z_k=-c$, $r_k=r$,
with $a$, $b$, $c$ positive and $r>0$.
Then, by \eq{eigeno31} we have
$r=ab-bc-ca >0$, or
\begin{equation*}
c< \frac{ab}{a+b},
\end{equation*}
while by \eq{eigeno31} and \eq{eigeno32}, again, we have
\begin{equation*}
(a+b-c)(ab-bc-ca)< -abc/r.
\end{equation*}
Thus, $a+b<c$.
Hence, $a+b < c < \frac{ab}{a+b}$, which implies
\begin{equation*}
(a+b)^2 < ab,
\end{equation*}
an impossibility, since $a$, $b$ and $c$ are non-zero positive numbers.
Since this is true for an arbitrary $k$, we conclude that even this case is disallowed.
\item \label{cas3}
One of the eigenvalues is positive and two negative. Again, let us fix an arbitrary $k$
and without loss of generality
assume that $x_k=a$, $y_k=-b$ and $z_k=-c$, with $a$, $b$, $c$ strictly positive
and $r_k=r\geq 1$.
Now, by \eq{eigeno31}, we have $bc -ab -ca=r>0$, implying
\begin{equation*}
\frac{a}{b}+\frac{a}{c}<1,
\end{equation*}
while \eq{eigeno31} and \eq{eigeno32} together imply
\begin{equation*}
(a-b-c)(bc-ab-ca)< abc/r.
\end{equation*}
Dividing both sides by $abc$ and rearranging the terms,
we have
\begin{equation*}
\frac{a}{b}+\frac{a}{c} +\frac{b+c}{a}> 5-\frac{1}{r},
\end{equation*}
where we used the inequality $b/c+c/b\geq 2$ for any pair of
positive definite numbers.
Thus,
\begin{equation*}
1+\frac{b+c}{a}> 5-\frac{1}{r},
\end{equation*}
leading to
\begin{equation*}
a< \frac{b+c}{4-1/r},
\end{equation*}
implying $a-b-c<0$.
Since this is true for all $k$ and $r_k\geq 1$,  we conclude that
$x_k+y_k+z_k=a-b-c<0$ for every $k$.
\item \label{cas4}
All the eigenvalues are negative. In this case, obviously, $x_k+y_k+z_k<0$.
\end{enumerate}
Thus, the eigenvalues of $G_k$ are either of the type in \ref{cas3} or \ref{cas4} if
$G_k$ are non-singular, with
the sum of eigenvalues, that is the trace, negative in both cases.

Finally, if $H_k$ is  singular for some $k$, then while we can not use \eq{go31},
the inequality \eq{go32} can still be used, as it does not involve an inverse of $H_k$.
From \eq{go32}, then, $\tr(G_k)< \det G_k=0$. We thus conclude that
the trace of the matrix $G_k$ is negative definite for any $k$,
whether $G_k$ is singular or not. This is in contradiction with the
equation
\begin{equation}
\sum_k\tr(G_k)=0. \label{eigeno33}
\end{equation}
obtained by taking trace on both sides of \eq{o3desum}.

For $\theta=\pi$ one can do the same analysis and obtain that the trace of the matrix $G_k$ is positive definite and so once again does not satisfy \eq{eigeno33}. Both the values of $\theta$ are not allowed simaltaneously as the supersymmetries preserved by them are mutually incompatible.

In the above analysis, we can incorporate ``wrapping numbers" \cite{Bianchi1}. The wrapping numbers are Jacobians of the
embedding of the six coordinates of the world volume of the D9-brane
onto $X$, which may be any non-zero positive or negative integer.
Inclusion of these factors is tantamount to multiplying both sides
of the inequality \eq{suseo32} by the sign corresponding to the
wrapping number of the $k$-th stack for each $k$. The expressions of
tadpoles will also be multiplied with the wrapping number of the
$k$-th stack. When the wrapping number is positive, that the above
argument goes through is obvious. For a negative wrapping number,
the inequality will change sign but so does its contribution to
tadpole and thus the above analysis remains valid. So in both cases
we find that the conclusion drawn above remains unaltered.

\subsubsection*{O9-plane}
Now we consider the orientifolding action with fixed 9-plane, which simply 
means the $Z_2$ consists of parity inverson only and is not combined with 
any space-time orbifold action like the other cases. 
We rewrite the relations \eq{suseo91}, \eq{suseo92}, \eq{o9tadp1} and \eq{o9tadp2}
respectively as
\begin{gather}
r_k > \tr ({G_k^{-1}})\det{G_k}, \label{go91}\\
\tr(G_k) = \frac{1}{r_k^2}\det {G_k},\label{go92}\\
\sum_k r_k \leq 16, \label{o9desum1}\\
\sum_k (G_k^{-1})\det{G_k} = 0. \label{09desum2}
\end{gather}
where we have used the strict positivity of $r_k$ and $\det\Omega$.
Moreover the form of the D9 tadpole contribution \eq{o9desum1} requires that we restrict to positive wrappings only.

Once again we rewrite equations \eq{go91} and \eq{go92} as, respectively,
\begin{gather}
r_k > x_k y_k+y_k z_k+ z_k x_k, \label{eigeno91}\\
r_k^2(x_k+y_k+z_k) = x_k y_k z_k,\label{eigeno92}\\
\sum_k ( x_k y_k+y_k z_k+ z_k x_k ) = 0, \label{eigeno93}
\end{gather}
and consider the various possibilities. Note that all the three relations remain unaltered if we flip the signs of all the eigenvalues simaltaneously. Therefore it is sufficient to consider two cases. The other possibilities can be obtained by flipping the signs of eigenvalues.

\begin{enumerate}

\item All the eigenvalues are positive. Let us choose $x_k = a, y_k=b, z_k=c$ for a particular stack with $(a,b,c)$ positive. \eq{eigeno91} and \eq{eigeno92} implies
\begin{gather}
\frac{1}{r_k^2} = \frac{a+b+c}{abc}, \label{eigeno94} \\
r_k > (ab +bc +ca). \label{eigeno95}
\end{gather}
The above two relations imply
\begin{gather*}
\frac{1}{r_k} > \frac{(a+b+c)(ab+bc+ca)}{abc} > 9 , 
\end{gather*}
where the second inequality follows from an argument
similar to the one already discussed in the case of O3-plane.
Since $r_k$ is a non-zero positive integer it cannot be less than $1/9$ and so this configuration cannot satisfy the supersymmetry requirements.
This also rules out the case where all eigenvalues are negative.

\item Two of the eigenvalues are positive and one negative. Let us choose $x_k = a, y_k = b, z_k = -c$ for a particular stack with $(a,b,c)$ positive. \eq{eigeno91} and \eq{eigeno92} imply
\begin{gather*}
r_k > ab -c(a+b) ,\quad\quad
a+b-c = - \frac{abc}{r_k^2} .
\end{gather*}
But then we have
\begin{gather*}
c - (a+b) = \frac{abc}{r_k^2} > 0,
\end{gather*} which implies $c > (a+b)$.
So this configuration can be compatible with the supersymmetry equations. Now we consider the condition \eq{eigeno93} which arises from tadpole condition. Since $c > (a+b)$ each summand corresponds to $k$-th stack in \eq{eigeno93}  
is of the form 
\begin{gather*}
ab - c(a+b) < ab - (a+b)^2 < 0.
\end{gather*}
Since all of them are negative
they cannot add up to zero. Thus
this configuration is not compatible with 
the tadpole cancellation condition.
This also rule out the case where two of the eigenvalues are negative as that can be obtained by flipping the signs of the eigenvalues which does not change the sign of tadpole.

\end{enumerate}

The conclusion of the analysis above is, therefore, that the
stabilization of K\"ahler moduli is impossible to achieve by
wrapping magnetized D9-branes corresponding to semi-homogeneous
vector bundles on the complex three-torus, for O9 and O3 orientifold planes. 
A similar general analysis for O5 and O7 orientifold planes turns out to be 
difficult because the D7 and D5 tadpoles as given in eqns. (\ref{o7tadp2}) and (\ref{o5tadp2}),
are not invariant under rotations along $T^6$ and therefore can not be analyzed in terms
of the eigenvalues such as $x_k, y_k, z_k$ of the matrices involving gauge fluxes.
Note, however, that in our analysis we have not assumed any restriction on the ranks or 
equality of the ranks of the
bundles. Therefore, our conclusions are valid for bundles of any rank. This, 
in particular, rules out stabilization with
Abelian fluxes, as has been suspected earlier \cite{Bianchi1}. It
has however been shown earlier \cite{kmr} that it may be possible to
stabilize all moduli, complex and K\"ahler, by considering
two-bundles and line bundles in conjunction. While the existence of
these bundles has not been rigorously established, these principal
bundles of rank two or higher seem to be the only possibilities for
these schemes with O3 or O9-planes to be successful. We now proceed to 
discuss the magnetized D7 brane systems, in order to prove the 
incompatibility of $T^6/Z_2^A$ and $T^6/Z_2^B$ compactifications
with moduli stabilization conditions in a supersymmetric theory.
\section{Seven-branes}
An alternative mechanism of stabilizing the K\"ahler and complex
structure moduli in a supersymmetric vacuum may be thought of, in
which space-filling D7-branes instead of D9-branes are wrapped on
holomorphic four-cycles of the complex three-dimensional compact
manifold. Such a scheme has been proposed with a real six-torus as
the compactification manifold \cite{akm2}, albeit in conjunction
with non-zero VEV of charged scalars. However, as mentioned above,
the different kinds of RR charges of the configuration must add up
to zero, for cancelation of tadpoles of all kinds. In this section
we demonstrate that even with space-filling D7-branes wrapped on
holomorphic four-cycles in the presence of O7 or O3-planes, D7-tadpole
cancelation and preservation of supersymmetry are mutually
exclusive. Therefore, a supersymmetric ground state is impossible to
realize even within this scheme. For completeness, We consider the constraints 
arising for O5 planes as well, while O9 case is trivially
ruled out.  O5 tadpole cancelation conditions are similar in structure
as the O7 and O5 examples of section-\ref{Semi-homogeneous bundles} and 
their discussions remain inconclusive for similar reasons. We now write down 
the consistency conditions and present the analyses.
\subsection{The constraints}
Let us consider orientifolded type--II string theory on the complex
three tori $X$ with a transverse $Op$-plane and space-filling magnetic
D7-branes wrapped on holomorphic four-cycles, $\Sigma$ of $X$
\cite{mmms}, holomorphicity being required by the preservation of
$\mathcal{N}=1$ supersymmetry. The K\"ahler form on $X$ is denoted
by $\Omega$, as before. The world-volume theory of the D7-brane is a
gauge theory and a generic configuration corresponds to a vector
bundle $E$ on the world-volume. The supersymmetry condition for such
a configuration is given by a non-linear generalization of hermitian
Yang-Mills equation \cite{Douglas:2001hw,mmms}:
\begin{gather} 
F_{\imath\bar\jmath}~ dz^\imath\wedge d{\bar z}^{\bar\jmath} = 0
\quad, \quad\quad \omega\wedge F = k (\text{vol}(\Sigma) -
\frac{1}{2}F\wedge F), \label{hym}
\end{gather} 
where $F$ denotes the
curvature associated with the bundle $E$ and $\omega$ denotes the
K\"ahler form on the four-cycle $\Sigma$ induced from $\Omega$.

In the presence of $Op$-plane, the supersymmetry imposes one more
condition. That the D-brane configuration preserves the same
supersymmetry as that of the $Op$-plane requires the
central charge
\begin{gather}
{\mathcal Z}= \int\limits_\Sigma e^{-i\omega}\ch(E),
\end{gather}
to satisfy the supersymmetry condition \eq{susy} where $\theta$ depends on the dimension of the orientifold plane. We enlist the supersymmetry conditions and tadpole cancellation conditions for different $Op$-planes in the following.

\subsubsection*{O3-plane}
In presence of O3-plane transverse to the compactification manifold $T^6$, the allowed values are either $\theta=0$ or $\theta=\pi$.
For $\theta=0$ D7-brane configuration needs to
satisfy the following constraints:
\begin{gather}
\omega \cdot F = 0 \quad ,\label{cchargeo31} \\
\int\limits_\Sigma [\frac{1}{2}\omega \cdot \omega - \ch_2(E) ] < 0. \label{cchargeo32}
\end{gather}
The first equation \eq{cchargeo31} implies that in presence of
O3-plane the value of $k=0$ in \eq{hym}.
For $\theta=\pi$ the preservation of supersymmetry requires,
\begin{gather}
\omega \cdot F = 0 \quad , \label{cchargeo71} \\
 \int\limits_\Sigma [
\frac{1}{2}\omega \cdot \omega - \ch_2(E) ] > 0. \label{cchargeo72}
\end{gather}
In addition we need to impose the vanishing tadpole condition. The vanishing of D7-brane
tadpole contribution requires the following integral summed over all
stacks,
\begin{gather}
\sum\limits_{\text stacks} \int d{\text vol}(\Sigma) {\text rk}(E) =
0, \label{d7tadpoleo31}\end{gather} evaluated on any four-cycle should
vanish. We have used ${\text rk}(E)$ for the rank of bundle $E$,
${\text vol}(\Sigma)$ for the volume-form of four cycle $\Sigma$. In
other words if we have D7-brane wrapped on some four-cycle $\Sigma$
whose volume takes a positive value to cancel this tadpole we need
to have another D-brane wrapped on some $\Sigma^\prime$ whose volume
takes negative value. This is equivalent to introducing wrapping numbers $n_k$ for $k$-th stack which takes both positive and negative values. We write down the condition in terms of wrapping numbers as
\begin{gather}
\sum\limits_k n_k ~ {\text rk}(E_k) = 0,\label{d7tadpoleo32}
\end{gather}
where $k$ denotes the stack and $n_k$ wrapping number for $k$-th stack.
Similarly the D3-brane tadpole contribution is
\begin{gather}
\sum\limits_k n_k ~ \ch_2(E_k) \leq 16 .\label{d3tadpoleo3}
\end{gather}

\subsubsection*{O7-plane}
In presence of O7-plane the only allowed value is $\theta=\pi$ and so the supersymmetry requires \eq{cchargeo71} and \eq{cchargeo72} have to be satisfied.
The tadpole cancellation condition depends on the wrapping of D7-brane with respect to the O7-plane. We will consider the case where D7-brane and O7-plane are on top of each other. That is the only configuration where one can cancel the D7-tadpole arising from O7-plane using D7-brane only. In that case, the vanishing condition for D7 and D3 tadpole are 
\begin{gather}
\sum\limits_k {\text rk}(E_k) \leq 16 , \label{d7tadpoleo7}\\
\sum\limits_k \ch_2(E_k) = 0 , \label{d3tadpoleo7}
\end{gather}
respectively.

\subsubsection*{O5-plane}
In these case, allowed values of $\theta$ are either $\theta=\frac{\pi}{2}$ or $\theta=-\frac{\pi}{2}$. For $\theta=\frac{\pi}{2}$ the supersymmetry condition \eq{susy} becomes
\begin{gather}
\omega \cdot F > 0 , \label{cchargeo51}\\   
\int\limits_\Sigma [
\frac{1}{2}\omega \cdot \omega - \ch_2(E) ] = 0. \label{cchargeo52} 
\end{gather}
For $\theta=-\frac{\pi}{2}$ the conditions become
\begin{gather}
\omega \cdot F < 0 , \label{cchargeo53}\\   
\int\limits_\Sigma [
\frac{1}{2}\omega \cdot \omega - \ch_2(E) ] = 0. \label{cchargeo54} 
\end{gather}
The tadpole cancellation condition depends on the orientifolding action.
For simplicity we assume that the D7-brane is wrapped on the $T^4$ spanned by the complex coordinates $\{z^1, z^2, {\bar z}^1, {\bar z}^2 \}$. The orientifolding action is given by $(z^2 , z^3, {\bar z}^2, {\bar z}^3 ) \longrightarrow (-z^2 , -z^3, -{\bar z}^2, -{\bar z}^3 )$ so that the O5-plane is wrapped on the $T^2$ spanned by the coordinates $\{ z^1 , {\bar z}^1 \}$. Then the D5-brane tadpole cancellation condition becomes
\begin{gather}
\sum\limits_k \ch_1(E_k)^{1{\bar 1}} \leq 16 , \quad\quad
\sum\limits_k \ch_1(E_k)^{\imath{\bar\jmath}} = 0 \quad \forall (\imath{\bar\jmath}) \neq (1 {\bar 1}) \label{d5tadpoleo5}.
\end{gather}  
Since the D7 brane does not generate any D9-brane charge the 
corresponding tadpole 
contribution is zero. The tadpole condition in the case of O5-plane is not 
invariant under a rotation along $T^6$ and therefore does not admit a general 
analysis.  
In what follows we will restrict ourselves to the general analysis for  O3 
and O7 orientifold planes.
In this scheme we leave out the O9-plane because this generates D9-brane tadpole. 
Clearly, one cannot cancel D9-brane tadpole using D7-branes only. 

Collecting all the conditions imposed by supersymmetry in
presence of O3 and O7-plane we get:
\begin{gather} F_{\imath\bar\jmath}~ dz^\imath\wedge d{\bar z}^{\bar\jmath} = 0,
\label{holomorphicity}\\
\omega\wedge F = 0. \label{susyeq}
\end{gather}
These two equations are the usual instanton equations and are common to both the O3 and O7 cases.
\eq{holomorphicity} implies the bundle $E$ is holomorphic. In order
to ensure that there is a solution of \eq{susyeq} one needs to show
that the bundle is stable. We will not get into the details of
stability criteria. However, in the following, we will use one of
the necessary conditions for the stability of the bundle which says
the discriminant of the bundle should be positive semi-definite.
In addition the supersymmetry requires either \eq{cchargeo32} or \eq{cchargeo72} has to be satisfied for O3 and \eq{cchargeo72} has to be satisfied for O7 orientifold planes. Moreover, one needs to cancel the tadpoles as well.
In the next subsection we will analyze all these constraints.

\subsection{Analysis of constraints}
\subsubsection*{O3-plane}
We begin with $\theta=0$. Let the closed
$(1,1)$-forms $C_I, I=1, 2, \cdots, h^{(11)}$ be an integer basis
for $H^{(1,1)}(\Sigma, \mathbf{Z})$ and $I_{IJ} = \int\limits_\Sigma
C_I\wedge C_J$ be the corresponding intersection matrix. Then in
cohomology we can expand
\begin{gather}
F = F^I C_I , \quad\quad \omega = \omega^I C_I.
\end{gather}
In this notation \eq{susyeq} and \eq{cchargeo32} become
\begin{gather}
(\omega \cdot F ) = 0, \quad (1/2)(\omega^2 - c_1^2) + c_2 < 0
\label{susyeqo31},
\end{gather} where we use the following notation to keep the
expressions concise. We write $I_{IJ}$ as the metric, $\omega \cdot
F = \omega^I I_{IJ} F^J$, $\omega^2 = \omega^I  I_{IJ} \omega^J$,
$c_1^2 =  c_1(E)^I I_{IJ} c_1(E)^J$ and $c_2 = \int\limits_\Sigma
C_2(E)$. For convenience we can make $I_{IJ}$ diagonal with positive
and negative entries. However, for a general four-manifold $\Sigma$
the number of positive entries $p$ in $I_{IJ}$ can be either $0$ or
$1$ \cite{barths} depending on whether $b_1$ of $\Sigma$ is even or
odd respectively.

We consider the two cases separately. If $p=0$ the metric
$I_{IJ}$ has signature $(-, -, \cdots, -)$. So there is no $\Sigma$
with $\omega^2 > 0$ on which D7-brane can be wrapped.

When $p=1$ the metric $I_{IJ}$ has signature $(+, -, \cdots, -)$.
For the Abelian case, {\it i.e.} when rank of $E$ is 1 \eq{susyeq}
reduces to
\begin{gather} \omega \cdot c_1 = 0, \quad
\omega^2 - c_1^2 < 0 \label{susyeqo32}.
\end{gather}
If $\omega$ is spacelike ( $\omega^2 > 0$) the first equation
implies $c_1$ is timelike but that in turn means $c_1^2 < 0$ and so
the second inequality cannot be satisfied.

When rank of $E$ is grater than 1  \eq{susyeqo31} reduces to
\begin{gather} \omega\cdot F = 0, \quad
(1/2)(\omega^2 - c_1^2) + c_2 < 0 \label{susyeqo33}.
\end{gather}
For our purpose, we can take the trace of first equation of
\eq{susyeqo33} and write $\omega\cdot c_1 = 0$. But a necessary
condition \cite{friedman} that the vector bundle $E$ of rank $r$ is
stable is the discriminant of $E$, which is given by
\begin{gather} \triangle =\frac{1}{2r^2} ( 2 r c_2 - (r-1) c_1^2 ),
\label{dis}\end{gather} is positive semi-definite. Eliminating $c_2$
from \eq{susyeqo33} and \eq{dis} we get
\begin{gather}
r \omega^2 - c_1^2 + 2 r^2 \triangle < 0 .
\end{gather}
Once again if $\omega$ is spacelike we have $c_1$ to be timelike and
so this inequality cannot be satisfied for a positive semi-definite
$\triangle$.

Thus we see for both the choices of $p$, supersymmetry requires
volume of $\Sigma$ is negative. On the other hand for tadpole
cancelation \eq{d7tadpoleo31} we need to introduce stacks where volume
of $\Sigma$ takes positive values as well. So tadpole cancelation
cannot be compatible with supersymmetry of O3-plane (and other
stacks).

For $\theta=\pi$ the analysis is similar except one would get only positive volume of $\Sigma$ and therefore it is not possible to 
satisfy \eq{d7tadpoleo31}.

\subsubsection*{O7-plane}
For O7-plane we consider equations \eq{susyeq} and \eq{cchargeo72} 
which, in this notation, become
\begin{gather}
(\omega \cdot F ) = 0, \quad (1/2)(\omega^2 - c_1^2) + c_2 > 0
\label{susyeq1o7}.
\end{gather}
In this case, $\omega^2$ has to be positive and so once again the first equation 
implies $(1/2)(c_1^2-c_2)$ is negative. This time it is compatible with 
\eq{cchargeo72}. However, the contribution to D3-brane tadpole 
charge \eq{d3tadpoleo7} is also $(1/2)(c_1^2-c_2)$.
Since for all the stacks this contribution is negative 
that cannot add up to zero. 
Therefore this configuration cannot satisfy both supersymmetry and vanishing 
of tadpole condition simultaneously.

\subsection{D5 branes} \label{D5-branes}
D5 branes are irrelevant, as far as the cancelations of tadpoles generated by
O9 and O7 planes, appearing in $Z_2^A$ and $Z_2^C$ orientifoldings
are concerned. In addition, it is known that an O3 plane tadpole contribution
can not be canceled by a magntized D5 in a supersymmetric way, thus ruling out
their use in $Z_2^B$ compactification as well. This leaves the last possibility,
namely the possible concellation of the D5 tadpoles generated by D9
and D7 banes in $T^6/Z_2^D$ orientifold.  However, as already stated earlier, 
a complete study of $Z_2^C$ and $Z_2^D$ orientifold of $T^6$ is left as a 
future exercise.

\section{Conclusions}

We conclude that in the schemes outlined above, in terms of worldvolume fluxes,
there is no supersymmetric ground state for a type I string compactification 
on $T^6$ as well as IIB orientifold compactifications on $T^6/\Omega (-)^{F_L} I_6$,
when magntized branes are used for generating D-term potentials for
closed string moduli. The result remains valid in the case of known 
closed
string fluxes relevant in the case of IIB on $T^6/\Omega (-)^{F_L} I_6$.
Indeed, these fluxes contribute to only D3 tadpoles with the same sign 
as ordinary D3 branes and therefore are not relevant for the
cancellations of unwanted tadpoles in this compactification 
using D9, D7 or D5 branes.
Other orientifold compactifications: $T^6/\Omega (-)^{F_L} I_2$
and $T^6/\Omega (-) I_4$, as we already mentioned earlier, 
do not admit a similar general analysis and so one has to check them 
individually depending on how the O-planes are positioned. However,
we have checked numerically for various possibilities and it 
turns out that it is unlikely to have a consistent solution to 
these systems. A general proof for these compactifications will be
useful and will be examined as a future exercise.
This suggests that we better look for supersymmetric ground state in
orbifolded orientifolds. One promising approach may be to consider orientifolds 
where the space-time orbifold part consists of shift symmetry, which reduce the 
number of twisted sector closed string moduli, or in particular 
completely eliminate them. One can then examine whether it is 
possible to build a 
realistic grand unified model with completely stabilized moduli\cite{akp}, 
using flux branes alone. This will provide an exact
CFT construction for moduli stabilization in a realistic 
setup and will therefore
be of great importance. We hope to return
to some of these issues in future.


\section*{Acknowledgement}
We benefitted  from discussions with Massimo Bianchi, Siddhartha Sen and Avijit
Mukherjee.

\end{document}